# Lifting, Loading, and Buckling in Conical Shells

Daniel Duffy,[1] Joselle M. McCracken,[2] Tayler S. Hebner,[2] Timothy J. White,[2] and John S. Biggins[1,*]

[1]*Department of Engineering, University of Cambridge, Trumpington Street, Cambridge CB2 1PZ, United Kingdom*

[2]*Department of Chemical and Biological Engineering, University of Colorado Boulder, 596 UCB, Boulder, Colorado 80309, USA*

(Received 23 March 2023; accepted 15 August 2023; published 3 October 2023)

Liquid crystal elastomer films that morph into cones are strikingly capable lifters. Thus motivated, we combine theory, numerics, and experiments to reexamine the load-bearing capacity of conical shells. We show that a cone squashed between frictionless surfaces buckles at a smaller load, even in scaling, than the classical Seide-Koiter result. Such buckling begins in a region of greatly amplified azimuthal compression generated in an outer boundary layer with oscillatory bend. Experimentally and numerically, buckling then grows subcritically over the full cone. We derive a new thin-limit formula for the critical load, $\propto t^{5/2}$, and validate it numerically. We also investigate deep postbuckling, finding further instabilities producing intricate states with multiple Pogorelov-type curved ridges arranged in concentric circles or Archimedean spirals. Finally, we investigate the forces exerted by such states, which limit lifting performance in active cones.

Liquid crystal elastomers (LCEs) are muscular actuating solids that contract uniaxially along their director on heating [1,2]. Flat LCE sheets containing concentric circle directors (+1 defects) correspondingly morph into conical shells [3,4] that can spectacularly lift thousands of times their own weight [5] [Fig. 1(a)]. Continuing the trend of soft materials reinvigorating shell mechanics [6–13], we investigate the buckling load of a conical shell as a fundamental limit on lifting performance.

In thin sheets, the prohibitive energetic cost of stretch ($\propto t$) relative to bend ($\propto t^3$) strongly favors almost-isometric deformations. Accordingly, an LCE cone's strength is usually attributed to the tip's singular Gauss curvature [3], which, via the *Theorema Egregium* [25,26], guarantees it cannot be flattened isometrically, naively suggesting that buckling requires a stretch-scale load, $f_* \propto t$. However, if bend were cost-free, the cone could buckle under zero load via tip inversion, which is isometric but requires a perfectly sharp ridge. Finite-threshold buckling therefore occurs via a short-wavelength mode where stretch and bend compete. Indeed, the classical result [15] predicts wavelengths $\propto \sqrt{t}$ and accordingly a stretch-bend load scaling $f_* \propto t^2$, similar to compressed cylinders [27] and pressurized spheres [28,29].

Here, we combine theory, numerics, and experiments to investigate the buckling and postbuckling of conical LCE shells. We find compressed cones deform predominantly in an outer boundary layer, which instigates buckling at much smaller loads than predicted classically. Shells are usually frustratingly weaker than their theoretical idealizations, but this is normally attributed to acute imperfection sensitivity [27,30]. Some previous works have also highlighted boundary conditions [13,31–41], but clarity on the key physics has not emerged. Cones provide a clear-cut, analytically tractable example where the boundary layer's influence is profound, even yielding a new thin-limit scaling: $f_* \propto t^{5/2}$.

We begin our investigation by using surface alignment to fabricate 30 μm thick LCE sheets with circular director patterns (following [16], Supplemental Material [14], Sec. S8A-B), which morph into cones with semiangle ≈60° on heating to 145 °C. Actuated cones were then squashed under a glass slide of controlled weight, and their deformations tracked with an optical profilometer (Supplemental Material [14], Sec. S9B). Corresponding numerics were conducted using MORPHOSHELL [42] to minimize a nonlinear shell energy for a cone squashed between frictionless slides. Both experimental and numerical cones buckle subcritically, popping into a noselike shape at a load far below that classically expected [Fig. 1(c)].

To investigate further, we consider a conical shell with semiangle $\alpha$ and radius $R$, and denote arc length along generators by $s$ and perpendicular distance to the cone axis by $r$. Compressing the shell by $\Delta h$ under a vertical force $f$ induces a membrane strain $\varepsilon$ and a bending strain $\beta = \kappa - \bar{\kappa}$, $\kappa$ and $\bar{\kappa}$ being the deformed and undeformed curvature tensors. The deformed state will then minimize the standard energy $E = \int W dA - f\Delta h$. For small strains (but large rotations) the appropriate energy density is [21]

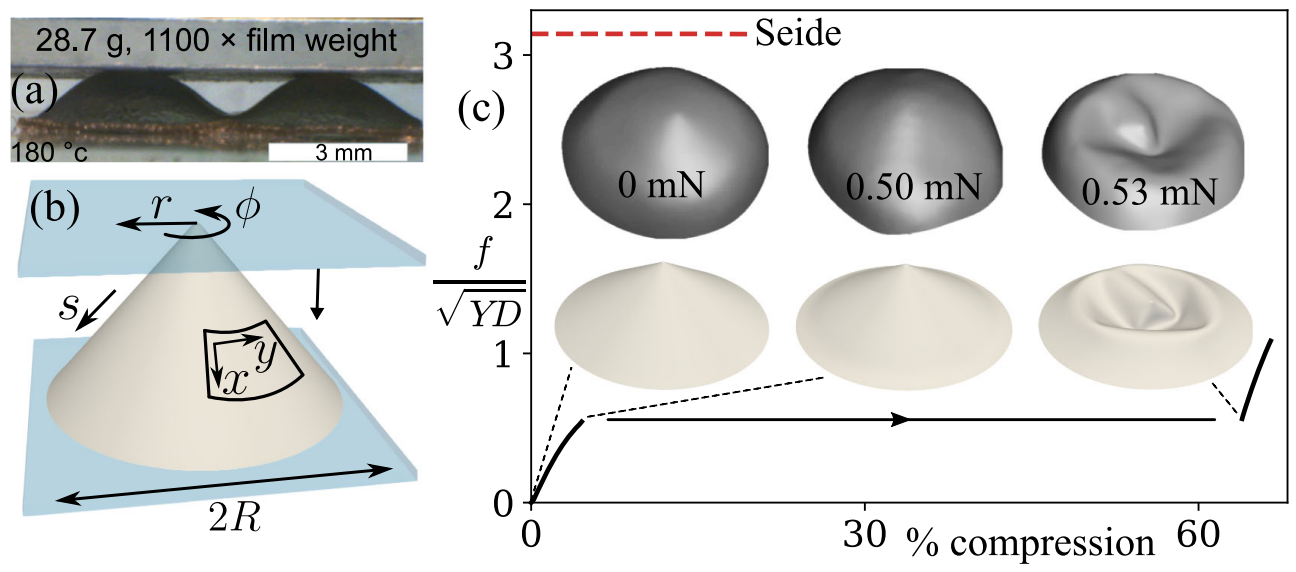

FIG. 1. (a) A $2 \times 2$ array of heat-activated LCE cones supporting a load $1100\times$ their own weight [5]. (b) Compressing a conical shell between frictionless slides. (c) Controlled-force compression of experimental (upper images) and numerical (lower images, force-compression plot) LCE cones ($\alpha \approx 60°$, $R = 200\ t$, $\nu = 1/2$), both exhibiting subcritical buckling. Using numerical and experimental (Supplemental Material [14], Sec. S9A) modulus values, both buckling thresholds are $\sim 20\%$ of the classical Seide threshold $\sim 3$ mN.

$$W(\varepsilon,\beta) = \frac{Y}{2(1-\nu^2)}Q(\varepsilon) + \frac{D}{2}Q(\beta), \qquad (1)$$

where $Y = \mathbb{E}t$ and $D = \mathbb{E}t^3/[12(1-\nu^2)]$ are the stretching and bending moduli for Young's modulus $\mathbb{E}$ and Poisson ratio $\nu$, while $Q(\tau) \equiv \nu \mathrm{tr}(\tau)^2 + (1-\nu)\mathrm{tr}(\tau\cdot\tau)$.

To probe the stability of an axisymmetric prebuckled base state, $\varepsilon_0$, $\kappa_0$, we consider small additional displacements, $\boldsymbol{u}$ tangentially and $w$ normally. Splitting the associated changes $\delta\varepsilon$ and $\delta\beta$ by order in displacement, retaining only terms important for short-wavelength perturbations gives [21]

$$\delta\varepsilon = \overbrace{\frac{1}{2}(\nabla\boldsymbol{u} + \nabla\boldsymbol{u}^T) + \kappa_0 w}^{\delta\varepsilon_1} + \overbrace{\frac{1}{2}(\nabla w \otimes \nabla w)}^{\delta\varepsilon_2},$$
$$\delta\beta = -\mathrm{Hessian}(w) = \delta\beta_1. \qquad (2)$$

The $\kappa_0 w$ strain is characteristic of shells, while the sole nonlinearity, $\delta\varepsilon_2$, allows inhomogeneous $w$ to relieve tangential compression: the basic mechanism of compressive buckling. Since the base state is an equilibrium, the leading energy change is quadratic in displacements:

$$\delta E = \int \{W(\delta\varepsilon_1, \delta\beta_1) + \mathrm{tr}[N(\varepsilon_0)\cdot\delta\varepsilon_2]\}dA, \qquad (3)$$

where $N(\varepsilon) = \partial W(\varepsilon,\beta)/\partial\varepsilon$ is the membrane stress.

In general, (3) involves covariant derivatives and integration over the entire curved shell. However, following Ref. [43], the anticipated short wavelength allows us to just consider a patch small compared to $r$ but large compared to wavelength. In this patch we may neglect *all* covariant considerations and use Cartesian coordinates $(x, y)$, aligning $x$ with $s$. Minimizing $\delta E$ variationally over displacements yields the expected tangential and normal force-balance equations, linear in displacements:

$$\nabla\cdot N(\delta\varepsilon_1) = 0, \qquad (4)$$

$$D\nabla^4 w + \mathrm{tr}(\kappa_0\cdot N(\delta\varepsilon_1)) - \nabla\cdot(N(\varepsilon_0)\cdot\nabla w) = 0. \qquad (5)$$

We satisfy (4) with a scalar Airy stress function $\psi$ such that $N(\delta\varepsilon_1) = (I\nabla^2 - \mathrm{Hessian})\psi \equiv \Lambda\psi$. Geometric compatibility of the strain then requires [43]

$$\nabla^4\psi - Y\,\mathrm{tr}(\Lambda(\kappa_0 w)) = 0. \qquad (6)$$

Traditionally one considers a membrane base state, with $\kappa_0 = \cos(\alpha)/r\,\hat{\boldsymbol{y}}\otimes\hat{\boldsymbol{y}}$ matching the undeformed cone, and $N(\varepsilon_0) = -f/(2\pi r\cos\alpha)\hat{\boldsymbol{x}}\otimes\hat{\boldsymbol{x}}$ following from vertical force balance. Both vary slowly over the cone, and so are effectively constant over the patch. We then search for oscillatory buckling solutions, substituting $(w,\psi) = (a,b)\exp[i(k_x x + k_y y)]$ into (5) and (6). As expected, the exponentials cancel, leaving algebraic equations that we solve for the ratio $a/b$ and buckling force $f$. Interestingly, force only depends on wave vector via $(k_x^2+k_y^2)^2/k_x^2 \equiv k_\circ^2$ and, minimizing over $k_\circ$, we find buckling commences at the classical Seide $f_* = 4\pi\sqrt{YD}\cos^2\alpha \propto t^2$ [15], with $k_\circ^4 = Y/(Dr^2)\cos^2\alpha$, corresponding to a "Koiter circle" of wave vectors with wavelengths $\propto \sqrt{rt}$, as is familiar from cylinders [27]. The threshold is radius independent, so buckling occurs over the entire cone simultaneously.

When squashing between slides we instead observe buckling at $\sim 20\%$ of Seide's value [Fig. 1(c)]. Moreover, the dominant prebuckling deformations are localized within a boundary layer of width $l \sim \sqrt{Rt}$ that is qualitatively different from the bulk's membrane state. Informatively, if the membrane state is artificially imposed at the boundaries, MORPHOSHELL reproduces Seide's threshold. We thus focus on this axisymmetric boundary layer, described by the local angle $\vartheta(s)$ [Fig. 2(a)] and the radial and vertical displacements $\Delta r(s), \Delta z(s)$, and with effectively constant $r \approx R$. Subtly, the rim's radial freedom allows near-complete $s$-stress relaxation, $N_{ss} \sim lN_{\phi\phi}/R$, consistent with stress equilibrium $N_{\phi\phi} = RN'_{ss}$. Consequently, $s$ strain is a pure Poisson effect of hoop strain, $\varepsilon_{ss} = -\nu\Delta r/R$, and the dominant balance between stretch and bend in (1) is simply

$$W \approx \frac{1}{2}Y(\Delta r/R)^2 + \frac{1}{2}D\vartheta'^2. \qquad (7)$$

Comparing these terms confirms the characteristic boundary-layer length scale, $l \equiv (R^2D/Y)^{1/4} \sim \sqrt{Rt}$. These terms and length scale also emerge from a conventional linear-elasticity treatment, but (7) also encompasses large rotations. Scaling all lengths by $l$ ($\mathsf{s} \equiv s/l$, $\Delta r(s) \equiv l\Delta r(\mathsf{s})$, etc.) then leads to the dimensionless boundary-layer energy:

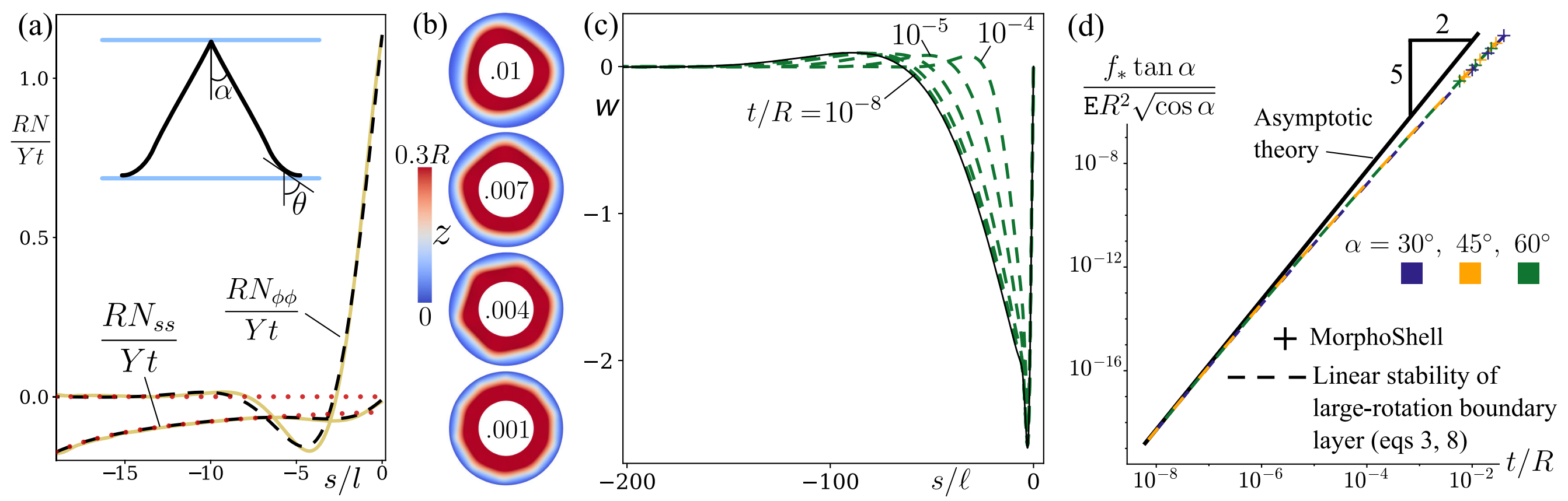

FIG. 2. (a) Stress profiles shortly before buckling for the cone in Fig. 1, with (beige) MORPHOSHELL and (dashed) boundary-layer (8) results agreeing well. The region of large compressive azimuthal stress near the boundary drives boundary-layer buckling at $f_* \propto t^2\sqrt{t/R}$, but is absent from the membrane base state (dotted). (b) Transitory MORPHOSHELL topographies just after boundary buckling, labelled by $t/R$, colored by elevation, with smaller thicknesses yielding larger azimuthal mode numbers. (c) Buckling mode shapes $w(s)$ from numerical linear stability analyses (dashed, $\alpha = 60°$) using our boundary-layer base state, converging to our thin-limit theory (solid) as $t/R$ decreases by factors of 10. (d) Buckling force against $t/R$ for $\nu = 1/2$ and various $\alpha$, with numerics converging to our thin-limit result (12).

$$\frac{E}{\pi l\sqrt{YD}} = \int \left[\Delta r^2 + \theta'^2 - \frac{f}{\pi\sqrt{YD}}\Delta z'\right] ds. \qquad (8)$$

The small-strain relations $\Delta r' = \sin\theta - \sin\alpha$ and $\Delta z' = \cos\alpha - \cos\theta$ allow us to write $E$ in terms of a single variable, $\Delta r$. Minimization via standard variational calculus then yields a nonlinear shape equation and boundary conditions (Supplemental Material [14], Sec. S1). Solving numerically using SciPy's solve_bvp [22] reveals a universal scaling form $\Delta r(s)$ for the boundary layer in all thin cones of given semiangle and dimensionless load $f/\sqrt{YD}$. The resultant shapes naturally include an outward flare, but oscillate into the cone, creating a band of inward displacement and large compressive hoop stress $N_{\phi\phi} = Y\Delta r/r$, which ultimately precipitates buckling. The smaller $N_{ss}$ follows from vertical force balance, and both shape and stresses agree well with full MORPHOSHELL simulations at realistic LCE thicknesses [Fig. 2(a)].

Equation (8) clarifies that geometrically large rotations in the boundary layer require $f \sim \sqrt{YD} \propto t^2$, the Seide buckling scale. In sufficiently thick simulated cones ($R \sim 10\ t$), a pronounced flared shape indeed forms under such loads before violent buckling eventually occurs via tip inversion (though accurate tip mechanics are beyond shell theory, so in practice other $f \sim \sqrt{YD}$ instabilities might occur instead, e.g., Seide buckling or "sleeve-rolling" boundary inversion).

However, instability in thinner cones (Fig. 1) is different, occurring at much smaller loads, and breaking azimuthal symmetry with a mode number $m$ that increases with $R/t$ [Fig. 2(b)], suggesting a competing instability with a higher thickness scaling. We thus use the incremental energy (3) to investigate the linear stability of the boundary-layer base state to azimuthally varying perturbations $w = lw(s)\cos(m\phi)$, and similarly for $\boldsymbol{u}$. Initially, we retain the full geometric nonlinearity of the boundary layer, requiring (3) to be understood covariantly on the base-state surface. Variational minimization then yields straightforward but cumbersome analogs of (4) and (5), and associated boundary conditions at the rim (Supplemental Material [14], Sec. S2), which we again solve using solve_bvp. We indeed uncover an instability with $w(s)$ localized to the rim and maximal in the compressive region, but with a surprisingly long decay into the cone over $\sim 100\ l$ [Fig. 2(c)]. The resulting $f_*$ values (minimized over $m$) agree well with MORPHOSHELL for realistic thicknesses, and present a clear thin asymptote to $f_* \propto t^2\sqrt{t/R}$ [Fig. 2(d)] with a correspondingly increasing azimuthal mode number and convergent (scaled) mode shape.

These numerical investigations clarify that, in the thin limit, $\Delta r$, $\Delta z$, and $\Delta\theta \equiv \theta - \alpha$ are all asymptotically small at buckling, suggesting a much simpler geometrically linear treatment of the base state will suffice. Linearizing the small-strain relations and substituting $\Delta r$ for $\theta$ in (7), we see that the small-amplitude boundary layer is better characterized by an $\alpha$-dependent length scale $\ell \equiv l\sqrt{2\ \sec\alpha}$. Using this modified length for nondimensionalization gives $W \propto 4\Delta r^2 + \Delta r''^2$, and hence, minimizing, the linear Euler-Lagrange equation $\Delta r'''' + 4\Delta r = 0$, as for a plate on a foundation. The solutions are $\Delta r \propto e^{\eta s}$ for $\eta^4 = -4$, whose oscillations produce the regions of hoop compression. Imposing the natural boundary conditions $\Delta r'' = 0$, $\pi\sqrt{YD}\Delta r''' = -f\sin\alpha$ and discarding the growing solutions gives the thin-limit base state

$$\Delta r = f e^{s} \cos(s) \sin(\alpha)/(2\pi\sqrt{YD}). \quad (9)$$

Reassuringly, this base state can be verified as the thin limit of a lengthy but routine linear elastic treatment (Supplemental Material [14], Sec. S3).

A further simplification arises because the base-state displacements are small, and localized within $\sim\ell$ of the rim. We may thus consider a patch at the rim, small compared to $r$ but large compared to $\ell$, and address stability with an Airy stress and the Cartesian equations (5) and (6). In the $(x, y)$ basis, the base state has $\kappa_0 = \mathrm{diag}(-\Delta\theta'/\ell, \cos\alpha/R)$ and $N(\varepsilon_0) = \mathrm{diag}(0, Y\ell\Delta r/R)$ to leading order which, unlike the classical case, vary over a scale $\ell$ and hence are inhomogeneous even within the patch. We find the equations take their simplest dimensionless form if we scale the force as $f \equiv 2\pi D f/(\ell \tan\alpha)$, and the fields as

$$(w, \psi) = (\ell w(s), Y\ell^3/R \cos(\alpha)\psi(s)) \cos(ky/\ell), \quad (10)$$

with $s = x/\ell$. This scaling retains many features of Seide buckling, including the scale of the wave vector, the natural stress scale, the relative sizes of $w$ and $\psi$, and the correspondingly small in-plane displacement $u \sim (\ell/R)w$. Remarkably, substituting into (5) and (6) gives equations that are not only dimensionless, but lack explicit dependence on $\alpha$ and $\nu$:

$$w'''' - 2k^2 w'' + k^4 w + 4\psi'' = 2fk^2 e^s(2\psi \sin s - w \cos s),$$
$$\psi'''' - 2k^2\psi'' + k^4\psi - w'' = -fk^2 w e^s \sin s. \quad (11)$$

Since the patch extends to the rim, we also require boundary conditions: To leading order, given the anticipated scalings, $w = 0$ (zero vertical displacement) and $w'' = \psi = \psi' = 0$ [natural conditions from varying (3)]. We again solve (11) with solve_bvp, sweeping through $k$ to find the mode that becomes unstable at the smallest $f$. The resultant mode shapes reproduce the thin limit of our previous approach [Fig. 2(c)], with first instability at $f_* = 62.7\ldots$, $k_* = 0.252\ldots$ (Supplemental Material [14], Fig. S4). Restoring dimensions yields the thin-limit buckling threshold

$$f_* = (278.4\ldots)(YD^3/R^2)^{1/4}\sqrt{\cos\alpha}\,\cot\alpha \propto t^{5/2}, \quad (12)$$

agreeing with Fig. 2(d), with azimuthal mode number $m_* = (0.178\ldots)(R^2Y/D)^{1/4}\sqrt{\cos\alpha} \propto \sqrt{R/t}$. Thus, larger cones are in fact weaker for a given $t$; conversely, smaller cones are stronger, though this is ultimately limited by the alternative $f_* \propto t^2$ modes noted earlier.

Figure 2(d) reveals that $f_*$ asymptotes remarkably slowly, perhaps due to the surprisingly long-ranged mode shape: a reminder to be cautious when exploiting "thinness" in shells. Although our experiments and MORPHOSHELL simulations ($R \approx 100t$) are preasymptotic, they nevertheless exhibit the same mechanical character: a region of compressive azimuthal boundary-layer stress initiates azimuthal buckling long before bulk instability. The cone's load-bearing capacity then drops drastically, and large postbuckling deformations immediately propagate deep into the bulk. The near collapse of curves for different $\alpha$ in Fig. 2(d) furthermore shows that, surprisingly, the $\alpha$ dependence of our asymptotic result pertains even far from the asymptote.

A compelling feature of soft solids is that buckling need not precipitate failure, allowing creative use of the resultant morphing [44–51]. In this spirit, we now squash cones far beyond their initial instability, using displacement control to explore full hysteretic cycles (Fig. 3). We consider cones formed by concentric-circle directors on both disks and squares, since both are used as LCE actuators [5,52]. Numerically, MORPHOSHELL finds multiple successive instabilities, producing striking shapes: disk-type cones yield increasing numbers of concentric circular ridges [Figs. 3(a) and 3(b)], evoking the exact isometries of a cone, with multiple sharp inversions, although blunted in the spirit of Pogorelov [53–55]. In squashing, subsequent ridges form via violent rim inversions, while in unsquashing they are annihilated centrally by tip inversion (Supplemental Material [14], Movies M1–3). Simulated square-type cones tend to instead exhibit spiral ridges, which (un)wind continuously in (un)squashing (Supplemental Material [14], Fig. S5, Movies M4–6). Experimentally, square-type samples were fabricated with a slightly different chemistry [16,23,24], and displacement control was implemented by using a digital caliper to control and measure the height of a slide (Supplemental Material [14], Sec. S8C and 9B). These cones switched between concentric-circle and spiral forms across successive load cycles [Fig. 3(c)], suggesting a delicate balance: a topic for future work.

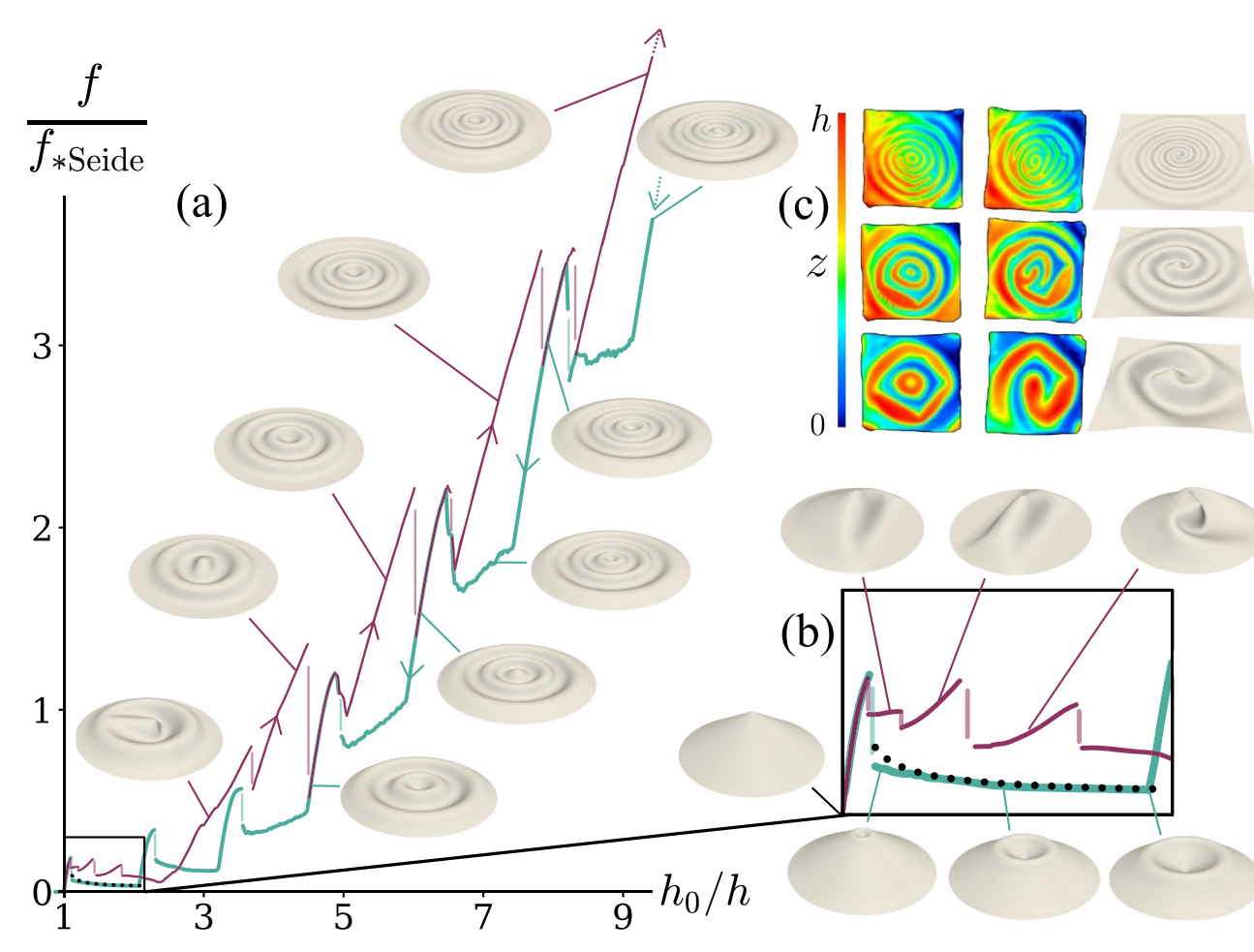

FIG. 3. (a) Vertical force against height ratio (initial/squashed) for an LCE cone compressed quasistatically in MORPHOSHELL ($\alpha = 60°$, $R = 100\ t$), with (un)squashing in (teal) purple. (b) Enlarged view of the small-compression region, where the unsquashing single-ridge force agrees well with our theory (dotted). (c) Height-colored experimental topographies of a square-type LCE cone, with two successive compressions of the same sample (left, middle) exhibiting both concentric-circle and spiral ridges, and (right) a spiral ridge in a similar simulated cone.

Unsquashing concentric ridges culminates in a single circular ridge moving toward the tip [Fig. 3(b)], yielding the cone's weakest states. The form and strength of such a low-force ridge can be understood by formulating an axisymmetric shape equation, as used to describe the boundary layer. Indeed, a sharp circular ridge would be an isometry, with zero stretch but divergent bend, leading instead to transversely blunted ridges, where stretch and bend compete [55]. The dominant energies are exactly those in (8), with blunting over the same scale $l$, except with ridge radius $\rho$ replacing $R$. The right-hand side of (8) is dimensionless so, substituting $l$, the energy of a ridge must be $E = 2g(\alpha)(YD^3\rho^2)^{1/4}$, where $g(\alpha)$ is a dimensionless geometric factor. Recognizing that the height of a ridge is $h = (R - \rho)\cot\alpha$, virtual work shows that the ridge exerts a force $f = -\partial E/\partial h = g(\alpha)\tan(\alpha)(YD^3/\rho^2)^{1/4} \sim t^2\sqrt{t/\rho}$. Interestingly, this force *decreases* with $\rho$, opposite to the spherical case [53].

We calculate $g(\alpha)$ for a ridge held between two distant circular clamps by again minimizing (8) numerically, confirming that ridges in steeper cones cost more energy (Supplemental Material [14], Fig. S6). The resultant force agrees well with MORPHOSHELL [Fig. 3(b)]. Interestingly this thickness scaling matches our asymptotic $f_*$, suggesting that, in scaling terms, boundary-layer buckling weakens the cone to the greatest possible extent. At sufficiently large radius a Pogorelov ridge in a sphere buckles into a polygon, under compressive azimuthal stress generated by the blunting deformations [7,54,56,57]: a cousin of boundary buckling. Cone ridges can also exhibit polygonality [Fig. 1(c)], but the presence of multiple ridges appears stabilizing, hence the circular ridges we observe at the deepest compressions.

A motivating question for LCE cones is what load they can lift, rather than merely support. Actuation starts with mild, weak cones, which will buckle into concentrically ridged states, and must unbuckle to lift. We therefore simulate a cone that activates from flat under a slide of fixed weight (initially supported by a small spacer so actuation commences), and explore how far it lifts. During unsquashing, ridges remain in contact with the slides, so the shell's cross section is zigzaglike [Fig. 4(a)]. Concentric ridges are thus equispaced, and spirals approximately Archimedean [Figs. 4(b) and 4(c)]. Since each ridge's force decreases with radius, the lowest-force state with $N$ ridges has all ridges at their largest possible radii. If the load exceeds this state's lifting force, the cone is stuck; otherwise it can lift all the way to the next such state. We thus predict a staircase of lifting heights as a function of weight. Assigning the single-ridge energy to each concentric ridge, a second virtual work argument (Supplemental Material [14], Sec. S7) yields a prediction for this staircase, in good agreement with MORPHOSHELL for small $N$ [Fig. 4(d)].

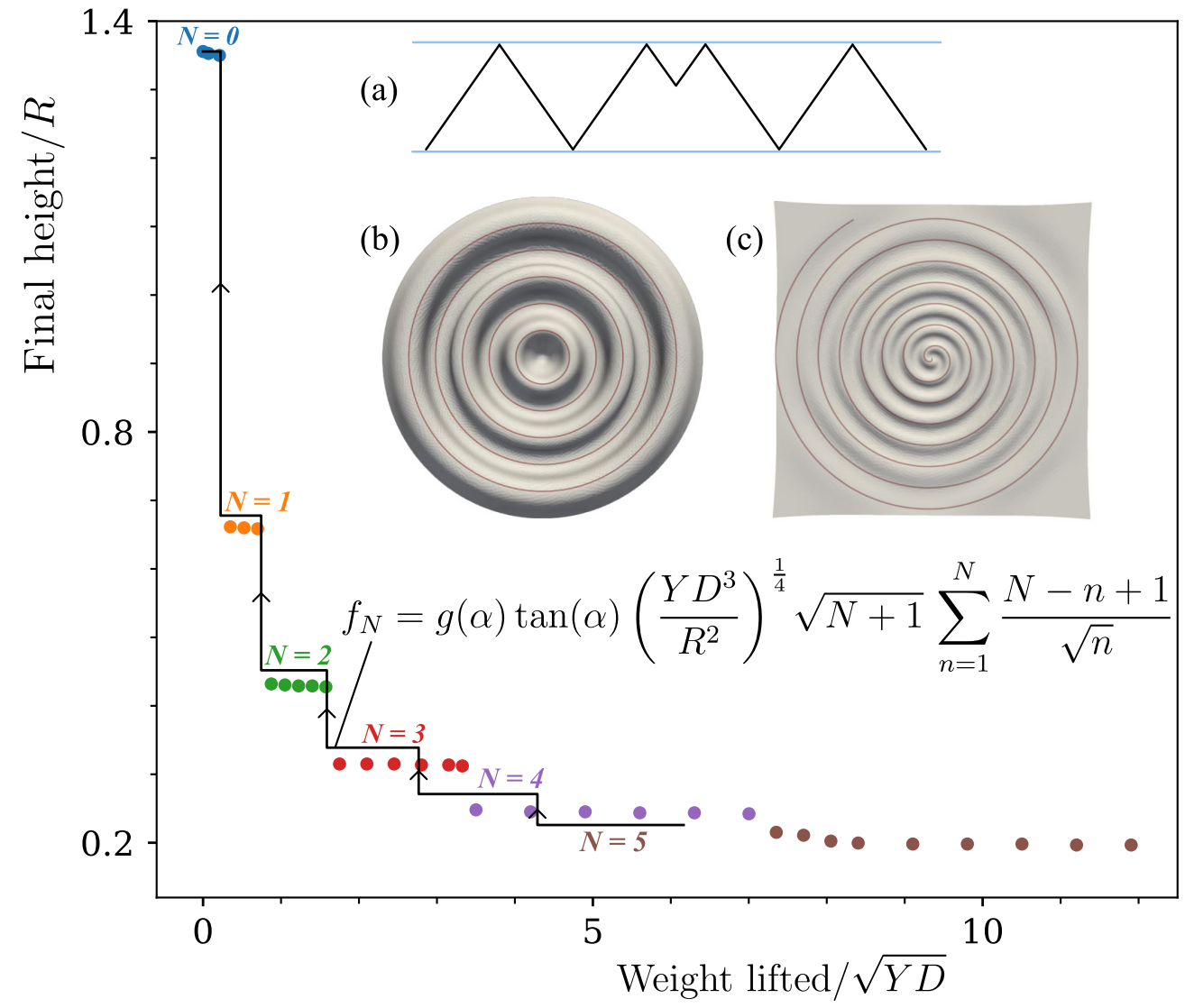

FIG. 4. (a) Profile sketch of a concentric-ridged lifter. (b) MORPHOSHELL confirms that concentric ridges are roughly equispaced while (c) spiral ridges are Archimedean. (d) Final height of a concentric-ridged lifter ($\alpha \approx 60°$, $R \approx 100\ t$) against weight lifted from a many-ridged state. Our simple theory (line) agrees well with MORPHOSHELL (markers) for small weights, but significantly underestimates lifting performance at large weights.

We conclude that only loads $\propto t^{5/2}$ can be lifted by LCE cones to large heights. This force is frustratingly close in scaling to the weak $\propto t^3$ forces offered by pure benders, and matches the scaling of boundary buckling. However, the unusual quantized height-load relationship allows for realization of discrete actuation strokes, which may be highly desirable in robust soft mechanisms or soft computation. At larger weights we observe increasing deviations from our staircase, with each ridge bearing load $\sim\sqrt{YD}$. This is unsurprising given the zigzag model must fail when ridge spacing and blunting become commensurate, but is good news for lifters: deeply ridged states can exert large forces, with potential for powerful small-stroke soft actuators.

Overall, our study reveals that, for frictionless boundary conditions, boundary-layer buckling limits the strength of cones, leading to a load-bearing capacity $\propto t^{5/2}$. This represents a significant knockdown from the classical result, *without* imperfections, and may contribute broadly to the practical weakness of shells. Active lifting to large heights via ridged states exhibits the same scaling. Future investigations will focus on using boundary conditions, geometry, and material parameters to optimize lifting performance and extract something closer to the full $\propto t$ energy budget of an actuating sheet.

J. S. B. was supported by a UKRI "future leaders fellowship" grant (Grant No. MR/S017186/1). D. D. was supported by the EPSRC Centre for Doctoral Training in Computational Methods for Materials Science (Grant No. EP/L015552/1). T. S. H. and T. J. W. acknowledge support from a Graduate Research Fellowship and DMR 2105369 from the National Science Foundation.

*Corresponding author: jsb56@cam.ac.uk